\documentclass[12pt]{JHEP3} 

\usepackage{epsfig}
\usepackage{amsmath}
\newcommand\fverb{\setbox\fverbbox=\hbox\bgroup\verb}
\newcommand\fverbdo{\egroup\medskip\noindent
			\fbox{\unhbox\fverbbox}\ }
\newcommand\fverbit{\egroup\item[\fbox{\unhbox\fverbbox}]}
\newbox\fverbbox

\newcommand \beq{\begin{eqnarray}}
\newcommand \eeq{\end{eqnarray}}

\title{Instanton-induced crossover in dense QCD}

\author{Naoki Yamamoto
\\
Department of Physics, The University of Tokyo, Tokyo 113-0033, Japan\\
	E-mail: \email{yamamoto@nt.phys.s.u-tokyo.ac.jp}}

\abstract{We study the properties of an instanton ensemble 
in three-flavor dense QCD which can be regarded 
as an instanton plasma weakly interacting by
exchanging the $\eta'$ mesons. Based on this description,
we explore the chiral phase transition induced by the 
instanton ensemble at high baryon density
in analogy with the Berezinskii-Kosterlitz-Thouless transition.
Using the renormalization group approach, we show that the instanton 
ensemble always behaves as a screened and unpaired plasma.
We also demonstrate that the chiral condensate in dense QCD 
is proportional to the instanton density.
}


\begin{document} 

\section{Introduction}
Topological excitations play crucial roles for 
understanding the properties of various systems 
in condensed matter physics and particle physics.
For example, $O(2)$ spin system in two-dimension is equivalent to a vortex 
ensemble interacting by 
two-dimensional Coulomb potential; it shows a second order phase 
transition from
a system composed of vortex dipoles to a vortex plasma as temperature 
increases.
This is known as the Berezinskii-Kosterlitz-Thouless phase transition 
\cite{BKT}.
Another remarkable example is the three-dimensional compact quantum 
electrodynamics (QED).
It can be described by an equivalent interacting magnetic monopole ensemble 
and shows a crossover as a function of the coupling constant $e$ 
\cite{P77}.
As a result, the area law of the Wilson loop, 
or the confinement of the fundamental charge, persists for
arbitrary value of $e$.

Moreover, the instanton ensemble have succeeded in illustrating many 
features of the vacuum of 
four-dimensional quantum chromodynamics (QCD) and its hadronic observables 
\cite{SS98}.
Most importantly, it provides a qualitative understanding of 
the spontaneous breaking of chiral symmetry in the QCD vacuum
as well as a possible mechanism of its restoration at finite temperature:
numerical calculations in the instanton liquid model 
show that the chiral restoration corresponds to
a transition from an unpaired instanton plasma at low temperature to 
instanton-antiinstanton molecules at high temperature 
in the physical case of up, down and strange quarks \cite{SS96}.

Recently, it was shown in the Ginzburg-Landau approach 
to three-flavor dense QCD \cite{HTYB06, YTHB07} that
 the interplay between the quark-antiquark pairing
(chiral condensate) and the quark-quark pairing (diquark condensate)
originating from the instanton-induced interaction may lead
to a smooth crossover between the hadronic phase and the 
color superconducting (CSC) phase \cite{CSC}. 
If such a crossover is realized, the coexistence phase of the chiral and 
diquark condensates extends to the region of high baryon density.
 However, the dynamical roles of the instanton ensemble  in 
 such a system  have not been  fully studied  in the  literatures
  except for a seminal work on the instanton description 
 of two-flavor color superconductivity (2SC) ~\cite{SSZ01}.
 It was shown in Ref. \cite{SSZ01} that the low-energy dynamics of 
 two-flavor dense QCD can be described by a nonideal instanton ensemble 
weakly interacting by exchanging the $\eta$ mesons
 due to the fact that the system of instantons is dilute and
 the $U(1)_A$ symmetry is asymptotically restored at high density.
 In such a case, the $\eta$ meson can be regarded as the lightest
  asymptotic Nambu-Goldstone (NG) boson.
 By rewriting the low-energy effective Lagrangian of the $\eta$ meson in 
the Coulomb gas representation
 via a duality mapping, two-flavor dense QCD reduces to an instanton 
ensemble where instantons (antiinstantons)
interact with each other by four-dimensional Coulomb potential generated by 
topological charges.

In the present paper, we will generalize the idea of 
Ref.~\cite{SSZ01} to three-flavor QCD: 
 We will first provide a complete derivation and its justification of the
instanton description of three-flavor dense QCD which was partially
 suggested  but was not fully explored in Ref.~\cite{SSZ01}.
 Then we will investigate the properties of the instanton ensemble
  using the renormalization group approach 
 and show that the instanton ensemble behaves as a screened and unpaired 
plasma. Thus, the chiral condensate inevitably exists 
  even at high baryon density regime.  This is consistent with 
  the previous finding in Refs.~\cite{HTYB06, YTHB07} and constitute
  a dynamical demonstration of the coexistence of the 
   chiral and diquark condensates at high density.

Throughout this paper, we will limit ourselves to three-flavor quark 
matter
with two light degenerate up and down quarks ($m_u=m_d=m_{ud}$) 
and a medium-heavy strange quark ($m_s > m_{ud}$) 
at zero temperature and at finite baryon density 
\footnote{We will not consider another possibility of the exotic state
called quarkyonic phase at high baryon density \cite{MP07}.}. 
 We remark here that the light $\eta'$ meson and 
the diluteness of instantons enable us to treat the 
instanton calculations under analytical control at high baryon density:
This is not the case in the vacuum and at finite temperature
where the assumption of the random instanton liquid  needs to be
 introduced  \cite{SS98}.

The paper is organized as follows. In Sec.~\ref{sec:inst},
 after
describing the instanton ensemble of three-flavor dense QCD,
we  derive analytical formulas for the instanton density, the topological 
susceptibility and a dense version of the Witten-Veneziano relation. 
In Sec.~\ref{sec:RG}, we show that the system of
instantons at high baryon density always behave as a screened and unpaired 
plasma by using the renormalization group approach.
Also we illustrate  that the chiral condensate induced by the instanton 
plasma is proportional to the instanton density.
Sec.~\ref{sec:discussion} is devoted to conclusion and summary.
In appendix.~\ref{sec:spectra}, we give the mass spectra of meson 
excitations at high baryon density.

\section{Instanton ensemble at high baryon density}
\label{sec:inst}
Let us consider how the low-energy dynamics in three-flavor dense QCD 
can be described by a nonideal instanton ensemble 
weakly interacting by exchanging the $\eta'$ mesons.
 Although the method employed in this section is
  motivated by the approach proposed in  Ref.~\cite{SSZ01}, 
  a complete derivation and its justification 
   for not-fully-explored three-flavor case is given here.
 First of all, owing to the inverse meson mass ordering, 
$m_{\eta'}<m_{K}<m_{\pi} < m_{\eta}$, which is caused by the explicit 
breaking of the flavor $SU(3)$ symmetry ($m_s > m_{ud}$) \cite{SS00}, 
we can focus on the low--energy effective Lagrangian of the $\eta'$ meson
at high baryon density. 
This ideal situation  has not been realized  in the two-flavor case,
because only two colors (red and green) participate in the
2SC pairing and there are not only asymptotically massless $\eta$ meson 
but unpaired (ungapped) blue quarks.

Our starting point is the three-flavor quark matter where the ground state 
is the 
color-flavor locking (CFL) color superconducting phase characterized by 
diquark condensates \cite{ARW}:
\beq
\label{eq:CFL}
\langle q_{Lb}^{j} C q_{Lc}^{k} \rangle = \epsilon_{abc} 
\epsilon_{ijk}[d_L^{\dag}]_{ai},
\nonumber \\
\langle q_{Rb}^{j} C q_{Rc}^{k} \rangle = \epsilon_{abc} 
\epsilon_{ijk}[d_R^{\dag}]_{ai}.
\eeq
Here $i,j,k$ ($a,b,c$) are flavor (color) indices and $C$ is the charge 
conjugation operator.
We define the $\eta'$ meson field $\phi$ as
\beq
d_L d_R^{\dag}=\left| d_L d_R^{\dag} \right|  e^{i\phi}.
\eeq
The field $\phi$ transforms as $\phi \rightarrow \phi + 4 \alpha_A$
under the $U(1)_A$ rotation $q_L \rightarrow e^{-i\alpha_A} q_L$.
The low-energy effective Lagrangian of the $\eta'$ meson at high density is 
given by \cite{MT00,SSZ,S02}:
\beq
\label{eq:eta'-Lagrangian}
{\cal L}&=& \frac{3}{4} f_{\eta'}^2 \left[(\partial_0 
\phi)^2-v^2(\partial_i \phi)^2 \right]
-V(\phi), 
\nonumber \\
V(\phi)&=&- a M \cos (\phi-\theta),
\eeq
where $f_{\eta'}$ is the decay constant of the $\eta'$ meson and
$v$ is the velocity originating from the absence of Lorentz invariance in 
medium.
$V(\phi)$ is the potential induced by one-instanton contribution,
$\sim {\rm Tr}_{ij}\left[\hat M_{ik} (d_L^{\dag} d_R)_{kj} \right]$ 
with the quark mass matrix $\hat M={\rm diag}(m_u, m_d, m_s)$
and ``Tr" is taken over flavor indices. 
$\theta$ is the theta-angle, $M$ is defined as $M = {\rm Tr} \hat M$ and
$a$ is a $\mu$-dependent parameter which we will explicitly calculate 
below.
We neglect the multi-instanton contributions to $V(\phi)$ 
since they are suppressed due to the diluteness of 
instantons at high baryon density.
It should be remarked that the term $\sim {\rm Tr}_{ij}\left[\hat M_{ik} 
(d_L^{\dag} d_R)_{kj} \right]$ 
generates not only the mass of the $\eta'$ meson
but also those of other pseudoscalar mesons ($\pi$, $K$ and $\eta$).
The contribution of the ${\cal O}(\hat M^2)$-term to 
Eq.~(\ref{eq:eta'-Lagrangian})
does not change our discussion basically and is neglected here for 
simplicity.
This will be considered in more detail in Appendix \ref{sec:spectra}.

At sufficiently large quark chemical potential compared with the typical 
scale of QCD,
$\mu \gg \Lambda_{\rm QCD}$, 
$f_{\eta'}$ and $v$ are found by matching to their microscopic values 
\cite{SS00}:
\beq
\label{eq:weak-coupling}
f_{\eta'}^2=\frac{3\mu^2}{8\pi^2}, \ \ v^2=\frac{1}{3}.
\eeq 

In order to obtain the explicit form of $V(\phi)$, 
let us start with the instanton-induced six-fermion interaction \cite{t76, 
SVZ80, SS98}:
\beq
\label{eq:inst-int}
{\cal L_{\rm inst}}&=&e^{i\theta}\int d \rho n(\rho) \frac{(2\pi \rho)^6 
\rho^3}{6N_c(N_c^2 -1)} 
\epsilon_{i_1 i_2 i_3} \epsilon_{j_1 j_2 j_3} \left[
\frac{2N_c+1}{2N_c+4}(\bar q _{Li_1}q_{Rj_1})(\bar q_{Li_2} q_{Rj_2})(\bar 
q_{Li_3} q_{Rj_3}) \right. \nonumber \\
& & 
\left. 
-\frac{3}{8(N_c+2)}(\bar q_{Li_1} q_{Rj_1})(\bar q_{Li_2}\sigma_{\mu \nu} 
q_{Rj_2})
(\bar q_{Li_3} \sigma_{\mu \nu} q_{Rj_3})+ (L \leftrightarrow R)
\right] + {\rm h.c.}.
\eeq
Here $\rho$ is the instanton size, $N_c$ is the number of colors, 
$i_{1,2,3}$ and $j_{1,2,3}$ are flavor indices and 
$\sigma_{\mu \nu}=\frac{i}{2}[\gamma_\mu, \gamma_\nu]$.
The instanton size distribution $n(\rho)$ is given by \cite{S82,SS98}
\beq
\label{eq:n(rho)}
n(\rho)=C_N \left(\frac{8\pi^2}{g^2} \right)^{2N_c} \rho^{-5} 
\exp\left(-\frac{8\pi^2}{g(\rho)^2} \right)e^{-N_f\mu^2 \rho^2},
\\
\label{eq:C_N}
C_N = \frac{0.466\exp(-1.679N_c)1.34^{N_f}}{(N_c-1)!(N_c-2)!},\\
\label{eq:b}
\frac{8\pi^2}{g(\rho)^2}=-b\log(\rho\Lambda_{\rm QCD}), \ \  b = 
\frac{11}{3}N_c-\frac{2}{3}N_f,
\eeq
where $N_f$ is the number of flavors.
Replacing one of $\bar q_L q_R$ with $\hat M$ in 
Eq.~(\ref{eq:inst-int}) 
and taking the expectation value with respect to the CFL ground state 
(\ref{eq:CFL}),
where
\beq
\label{eq:d}
|d_L|=|d_R|=\sqrt{\frac{6N_c}{N_c+1}}\frac{\mu^2 \Delta}{\pi g},
\eeq
with $\Delta$ being the superconducting gap near the Fermi surface,
one finds \cite{MT00,SSZ,S02}:
\beq
\label{eq:V}
V(\phi)=- \int d\rho n(\rho) \frac{2(2\pi \rho)^4 \rho^3}{N_c(N_c-1)} 
 |d_L|^2 2M \cos(\phi-\theta).
\eeq
The integration over the instanton size $\rho$ above results in 
the form of Eq.~(\ref{eq:eta'-Lagrangian}), where the coefficient $a$
is given by
\beq
\label{eq:a}
a(\mu) = 
\frac{24}{N_c^2-1}C_N N_f^{-\frac{b+3}{2}} \Gamma\left(\frac{b+3}{2}\right) 
\left(\frac{8\pi^2}{g^2}\right)^{2N_c+1} 
\left(\frac{\Lambda_{\rm QCD}}{\mu}\right)^{b} \mu \Delta^2 , 
\eeq
with $\Gamma(x)$ being the gamma function.
The well-known infrared divergence in instanton calculation in the QCD 
vacuum 
is not seen here since the instanton screening factor, $e^{-N_f\mu^2 
\rho^2}$ 
in Eq.~(\ref{eq:n(rho)}) \cite{S82}, 
gives the small size $\rho \sim \mu^{-1}$ of 
instantons 
and regulate the integral.

By rescaling $\phi \rightarrow 2\phi/( \sqrt{3v} f_{\eta'} )$ 
and using the new coordinate $x_0=v\tau$ with the imaginary time $\tau$,
the effective action of $\eta'$ in Eq.~(\ref{eq:eta'-Lagrangian}) reduces 
to 
the Euclidean invariant form:
\beq
\label{eq:eta'-action}
S_E&=&\int d^4x[(\partial \phi)^2 - \lambda \cos \alpha(\phi - \theta)],
\\
\label{eq:lambda}
\lambda&=& \frac{a}{v}M, \ \ \alpha=\frac{2}{\sqrt{3v} f_{\eta'}}.
\eeq
We note that the parameter 
$\alpha$ is a function of chemical potential since $f_{\eta'} \sim \mu$.
The instanton potential gives the $\eta'$ mass as
\beq
\label{eq:eta'-mass}
m_{\eta'}^2= {\frac{\lambda}{2}}\alpha^2
=\frac{16 \pi^2 a}{3\mu^2}M,
\eeq
where the second equation holds from the weak coupling relation, 
Eq.~(\ref{eq:weak-coupling}).
Therefore, $a \rightarrow 0$ as $\mu \rightarrow \infty$ from 
Eq.~(\ref{eq:a}) and the $\eta'$ meson
is a NG boson at high baryon density limit.

Via a dual transformation,
the partition function for the action in Eq.~(\ref{eq:eta'-action})
reduces to the following form \cite{SSZ01}:
\beq
\label{eq:part-func}
Z 
&=& \int {\cal D} \phi e^{-S_E}
=  \int {\cal D} \phi e^{-\int d^4x (\partial \phi)^2}
e^{\lambda\int d^4x \cos \alpha \left(\phi(x)-\theta \right)}
 \nonumber \\
&=& \sum_{N_{\pm}=0}^{\infty}\frac{(\lambda/2)^{N}}{N_+ ! N_- !}
 \int d^4x_1 ... \int d^4x_{N} 
 \int {\cal D} \phi e^{-\int d^4x (\partial \phi)^2}
e^{i\sum_{i=0}^N Q_i \alpha \left(\phi (x_i)- \theta \right)},
\eeq
where the sum is taken over possible sets of $N_+$ ($N_-$) with positive 
(negative) charge $Q_i=\pm1$
located at the position $x_i$. In deriving the second line in 
Eq.~(\ref{eq:part-func}), 
we have used the relation,
\beq
\lambda \cos \alpha \left(\phi(x)-\theta \right)=
\frac{\lambda}{2}\sum_{Q=\pm 1}e^{iQ\alpha \left(\phi(x)-\theta \right)}.
\eeq
Integrating over the variable $\phi(x)$ in Eq.~(\ref{eq:part-func}), one 
ends up with \cite{SSZ01}
\beq
\label{eq:part-func2}
Z&=& \sum_{N_{\pm}=0}^{\infty} \frac{(\lambda/2)^{N}}{N_+ ! N_- !}
\int d^4x_1 ... \int d^4x_{N} 
e^{-i\theta \sum_{i=0}^{N}Q_i}e^{-\sum_{i>j=0}^{N}Q_i Q_j G(x_i-x_j)},
\eeq
which is a Coulomb gas representation of the original sine-Gordon model.
Since the $\theta$-angle is conjugate to the topological charge in QCD, 
$Q=\sum_i Q_i=N_+ - N_-$ is identified with the total topological charge 
and
$N_+$ ($N_-$) with the number of instantons (antiinstantons).
Also,
\beq
G(x_i-x_j)=\frac{\alpha^2}{8\pi^2 (x_i-x_j)^2}
\eeq 
is the four-dimensional Coulomb potential between instantons 
(antiinstantons).
Therefore, Eq.~(\ref{eq:part-func2}) exhibits that 
this system is an instanton ensemble in which instantons and antiinstantons 
with topological charge $Q_i=\pm 1$ interact with each other by the 
potential $G(x_i-x_j)$.

Note that we can treat our instanton calculations under completely
analytical control depending on two distinctive facts in dense QCD:

(i) Instantons are sufficiently dilute indicated by the parameter 
$\Lambda_{\rm QCD}/\mu \ll 1$,
which enables us to deal with the effects of instantons as a perturbation. 

(ii) The inverse mass ordering of pseudoscalar mesons, 
$m_{\eta'}<m_{K}<m_{\pi} < m_{\eta}$ \cite{SS00},
guarantees that the low-energy dynamics is dominated by the $\eta'$ mesons 
\footnote
{We neglect the exact massless $H$ boson associated with the breaking of 
the $U(1)_B$ symmetry 
since its dynamics is totally independent here and decouples 
from the low-energy effective Lagrangian of $\eta'$.}.
This is the characteristics with three-flavor and can be confirmed 
at sufficiently large baryon density
(See Eqs.~(\ref{eq:pi})-(\ref{eq:eta}) in Appendix~\ref{sec:spectra}).

More quantitative estimate on the domain of applicability of this instanton 
description 
will be discussed in Sec.~\ref{sec:discussion}.

\subsection{Instanton density and topological susceptibility}
\label{sec:prop}
In this subsection, we calculate quantities based on the instanton ensemble 
discussed above.

Multiplying $N_+$ ($N_-$) in the right hand side of 
Eq.~(\ref{eq:part-func}), 
one finds the expectation value of the instanton (antiinstanton) number as
\beq
\label{eq:1-mom}
\langle N_+ \rangle = \langle N_- \rangle = \frac{\lambda v}{2} V_4,
\eeq
with four-volume $V_4=\int d \tau d^3 x=\int dx_0 d^3x/v$.
This shows that the average of the topological charge 
$\langle Q \rangle=\langle N_+ \rangle - \langle N_- \rangle$ vanishes
and the instanton density as defined below reads
\beq
n_{\rm inst}=\frac{\langle N \rangle}{V_4} = \lambda v
\eeq
with $N=N_+ + N_-$.  This result has been obtained in Ref.~\cite{SSZ01}. 

Moreover, by using Eq.~(\ref{eq:part-func}), we generally obtain
the mixed factorial moments:
\beq
\label{eq:2-mom}
\left\langle \frac{N_+ !}{(N_+ - k)!}\frac{N_- !}{(N_- - l)!}   
\right\rangle
=\left( \frac{\lambda v}{2} V_4 \right)^{k+l},
\eeq
for arbitrary nonnegative integers $k$ and $l$.
Eq.~(\ref{eq:2-mom}) implies that instantons and antiinstantons 
independently follow the Poisson distribution,
\beq
f(x)={\rm e}^{-\beta}\frac{\beta^x}{x!},
\eeq 
with $\beta=\lambda v V_4/2=\langle N \rangle/2$, from the fact that
the $n$-th factorial moment of the Poisson distribution 
is equal to $\beta^n$.
This Poissonian behavior is usually assumed in the QCD vacuum \cite{SS98},
but it can be justified at high baryon density for a dilute system 
of interacting instantons and antiinstantons as anticipated.
Also, for the topological susceptibility defined by 
\beq
\chi_{\rm top}=\frac{\langle Q^2 \rangle}{V_4},
\eeq
we have a simple relation,
\beq
\label{eq:top_suscept}
\chi_{\rm top}=n_{\rm inst}=\lambda v,
\eeq
as a property of the Poisson distribution.
By the use of Eq.~(\ref{eq:top_suscept}), 
the $\eta'$ mass in Eq.~(\ref{eq:eta'-mass}) reduces to
\footnote
{The topological susceptibility in Eq.~(\ref{eq:top_suscept})
and the Witten-Veneziano relation (\ref{eq:witten}) 
are consistent with the results of Ref.~\cite{S02}
at high baryon density where the two-instanton term is negligible,
though the factor $v^2$ in Eq.~(\ref{eq:witten}) does not appear in \cite{S02}.
This difference comes from the fact that our $\eta'$ mass is defined 
to satisfy the dispersion relation $E^2=v^2({\bf p}^2 + m_{\eta'}^2)$ 
while that in Ref.~\cite{S02} 
is the pole mass, i.e., the energy of $\eta'$ at ${\bf p}={\bf 0}$, 
$m^{\rm (pole)}_{\eta'}=v m_{\eta'}$.}
\beq
\label{eq:witten}
m_{\eta'}^2=\frac{2\chi_{\rm top}}{3f_{\eta'}^2 v^2}.
\eeq
This is a dense version of the Witten-Veneziano relation \cite{WV79} 
obtained as a natural application of the instanton ensemble, 
which is not given in Ref.~\cite{SSZ01}.

\section{Renormalization group analysis on instanton ensemble}
\label{sec:RG}
In this section, we consider the possible phases of instantons at high 
baryon density
on the basis of the instanton description in Sec.~\ref{sec:inst}.
In the following, we set $\theta=0$ for simplicity 
since all the arguments are independent of the parameter $\theta$.
In order to explore and compare the general properties of phase transitions 
induced by the $D$-dimensional 
topological excitations 
($D=2$ for vortices, $D=3$ for monopoles and $D=4$ for instantons),
we generalize Eq.~(\ref{eq:eta'-action}) to the $D$-dimensional sine-Gordon 
model whose action is given by
\beq
\label{eq:sine-Gordon}
S_D=\int d^D x \left[(\partial \phi)^2 - \lambda_D \cos \alpha \phi \right].
\eeq
Here $\alpha$ is the parameter with the mass dimension $1-D/2$,
which is introduced after appropriate rescaling the field $\phi$ 
so that we normalize the coefficient of the kinetic term to be 1.

The long-range Coulomb force between topological excitations requires
effects of many-body dynamics or quantum fluctuations. 
For this purpose, we shall now perform the Wilson renormalization group 
(RG) approach and divide $\phi(x)$ into two components, $\phi=\phi'+ \delta \phi$ 
with low-momentum part $0<k<\Lambda'$ and high-momentum part 
$\Lambda'<k<\Lambda$ respectively,
where $\Lambda'$ is smaller than $\Lambda$ by an exponential factor.
This RG analysis for $D=2$ has been already carried out in Ref.~\cite{BKT}, 
and we extend 
it to the case of $D \ge 3$ in the following.
Considering how the small coupling $\lambda \ll 1$ with the predominantly 
Gaussian fluctuations
shifts after the RG transformation, the change of the potential term can be 
calculated by 
integrating out the momentum shell $\Lambda'<k<\Lambda$ as
\beq
\langle \cos \alpha (\phi'+\delta \phi) \rangle 
&=& \frac{1}{2}\left(e^{i \alpha \phi'}e^{-\alpha^2\langle \delta \phi^2 
\rangle_D /2} + {\rm c.c.}  \right),
\eeq
with
\beq
\label{eq:integral}
\langle \delta \phi ^2 \rangle_D
=\sum_{\Vec{k}} \frac{1}{{\Vec{k}}^2}
&=& \int \frac {d \Omega_D}{(2\pi)^D} 
\int_{\Lambda'}^{\Lambda}\frac{k^{D-1}}{k^2}dk,
\eeq
where $\Omega_D$ is the surface area of a unit sphere in Euclidean 
$D$-dimension. 
Therefore, the form of sine-Gordon action, Eq.~(\ref{eq:sine-Gordon}), is 
preserved and changed to
\beq
S_D \rightarrow {S_D}' 
= \int d^D x \left((\partial \phi')^2 -\lambda_D^* \cos \alpha \phi' 
\right),
\eeq
with the coupling constant
\beq
\label{eq:coupling}
\lambda_{D=2}^* &=& 
x^{\alpha^2 / 4\pi}\lambda_D, \nonumber \\
\lambda_{D \ge 3}^* &=&
{\rm exp}\left[\frac{\alpha^2\Lambda^{D-2}(x^{D-2}-1)}{(D-2)2^D 
\pi^{D/2}\Gamma(D/2)}  \right] \lambda_{D}, 
\eeq
for $D=2$ and $D \ge 3$ respectively.
Here we define the renormalization scale $x=\Lambda'/\Lambda<1$.
At the same time, the kinetic term $(\partial \phi')^2$ is effectively 
reduced
by the factor of $x^2<1$ independent of the dimension $D$, 
since $\partial_{\mu}$ is of order $\Lambda$ and $\phi'$ is of order 
$\Lambda'$.

The systems described by the $D$-dimensional ($D=2,3,4$) sine-Gordon model 
are summarized as follows:
(a) two-dimensional $O(2)$ spin model with the nearest neighboring 
interaction $J$ \cite{BKT}, 
(b) three-dimensional compact QED with the coupling constant $e$ \cite{P77},
and
(c) four-dimensional dense QCD with quark chemical potential $\mu$. 
They are respectively equivalent to an ensemble of 
vortices, magnetic monopoles and instantons interacting by 
the $D$-dimensional Coulomb potential.
The resultant orders of phase transitions are summarized in 
Table.~\ref{tab}.
The parameter $\alpha$ in each case is also given.

\TABLE[h]
{\begin{tabular}{|c|c|c|c|c|}
\hline 
     & D & topological excitation & parameter $\alpha$     & order of phase 
transition \\ \hline \hline
(a)  & 2         & vortex              & $\alpha \propto  \sqrt{J/T}$   & second 
order  \\ 
(b)  & 3         & magnetic monopole   & $\alpha \propto  1/e$           & 
crossover  \\ 
(c)  & 4         & instanton           & $\alpha = 2/(\sqrt{3v} f_{\eta'}) 
\sim 1/\mu$  & crossover  \\
\hline
\end{tabular}
\caption{: Order of phase transitions of 
(a) two-dimensional $O(2)$ spin model with the nearest neighboring 
interaction $J$ \cite{BKT},
(b) three-dimensional compact QED with the coupling constant $e$ \cite{P77} and
(c) four-dimensional dense QCD with quark chemical potential $\mu$.
In each case, $D$-dimensional ($D=2,3,4$) sine-Gordon model is 
equivalent to an ensemble of Coulomb-like interacting topological 
excitations.
Parameters $\alpha$ to exhibit phase transitions are also shown.}}
\label{tab}

\subsection{Case (a): two-dimensional $O(2)$ spin model }
As a pedagogical demonstration,
we first recall the case (a) and consider which is overwhelming after the 
RG transformation, 
the potential term or the kinetic term in accordance with Ref.~\cite{BKT}.
From Eq.~(\ref{eq:coupling}),
when $\alpha^2/4\pi>2$, the potential term is suppressed by fluctuations
so quickly that it is irrelevant compared to the kinetic term.
Therefore, the system can be described only by the spin wave in this case. 
In the language of the Coulomb gas representation, this corresponds to an 
insulating phase where vortex and antivortex occur in pairs. 
Otherwise, i.e., $\alpha^2/4\pi<2$, 
the potential term takes over the kinetic term regardless of the initial 
value of $\lambda$
and the system is locked in one of the cosine minima $\phi=2\pi n/\alpha$ 
with integer $n$.
This corresponds to a plasma phase where the Coulomb potential is screened 
by the free vortices.
As a result, the system changes from vortex dipoles to a vortex plasma
on reaching $\alpha^2/8\pi=1$ as temperature increases.
Also we can easily check that the transition temperature 
$T_{\rm c}={J}/{8\pi}$
is identical to the prediction 
obtained from the interplay between the free-energy and the entropy of the 
vortex ensemble \cite{BKT}. 

\subsection{Case (b): three-dimensional compact QED }
\label{sec:b}
For $D \ge 3$, on the other hand, 
the kinetic term $\sim x^2$ vanishes while the potential term $\lambda$ 
remains finite 
in the limit $x \rightarrow 0$, unlike $\lambda$ also vanishes for $D=2$.
This originates from the fact that the integral in Eq.~(\ref{eq:integral}) 
is infrared divergent only for $D=2$,
but is finite for $D \ge 3$.
Therefore, the kinetic term is more suppressed than the potential after the 
RG transformation
and topological excitations for $D \ge 3$ always behave as a screened and 
unpaired plasma.

As a result, in the case (b), the magnetic monopoles resides in a screened 
plasma phase 
and show a crossover as a function of the coupling constant $e$.
Since the area law of the Wilson loop can be proven for the strong coupling 
limit $e \gg 1$,
it leads to a well-known conclusion that 
the confinement of the fundamental charge 
persists for arbitrary value of $e$ in the three-dimensional compact QED, 
which was first shown in Ref.~\cite{P77}.

\subsection{Case (c): four-dimensional QCD at finite baryon density}
Let us now turn back to the pending question of our interest,
whether the system of instantons acts as an instanton plasma or 
they couple into molecules in the case (c).
In an analogous fashion to the previous subsection, 
we find that the system of instantons always behaves as a screened and 
unpaired plasma and shows a crossover as a function of $f_{\eta'} \sim \mu$.
Since unpaired instantons induce the formation of quark-antiquark 
pairing and give nonvanishing chiral condensate,
our result implies that the chiral condensate will remain finite 
in the region of high baryon density
\footnote{
The application of our argument here to two-flavor QCD is not 
straightforward, since there are not only light $\eta$ mesons
but nearly massless unpaired blue quarks in the 2SC phase, 
as mentioned in Sec.~\ref{sec:inst}.
However, the instanton liquid model with two-flavor shows 
a tendency towards chiral restoration by forming instanton
molecules at high baryon density \cite{S98}. 
}.

More quantitatively, we can calculate the chiral condensate in relation to 
our instanton ensemble.
The minimum of the potential $V(\phi)$ in Eq.~(\ref{eq:eta'-Lagrangian}) is 
given at $\phi=\theta$:
\beq
V(\phi)_{\rm min}=-aM.
\eeq
Differentiating this energy with respect to $M$, one obtains the chiral 
condensate as
\beq
\label{eq:chiral}
\langle \bar q q \rangle_{\rm csc} 
= -a
= -\frac{n_{\rm inst}}{M},
\eeq
where we have used Eqs.~(\ref{eq:lambda}) and (\ref{eq:top_suscept}).
Eq.~(\ref{eq:chiral}) is a novel relation 
connecting the chiral condensate to the instanton density 
in dense QCD.
Since the instanton density rapidly decreases at high baryon density 
like $n_{\rm inst} \propto \lambda \sim \mu^{1-b}$ with $b=9$ for 
$N_c=N_f=3$ from Eqs.~(\ref{eq:b}) and (\ref{eq:a}),
the chiral condensate is highly suppressed
(but remains finite) like $\langle \bar qq \rangle_{\rm csc} \sim \mu^{1-b}$. 

This is a remarkable consequence, since previous studies using three-flavor
effective model calculations
such as the Nambu-Jona-Lasinio (NJL) model \cite{B05} and the random matrix 
model \cite{VJ05},
exhibit the pure CSC phase without the chiral condensate is realized at 
high baryon density.
This difference comes from the fact that they neglect the effects of 
instantons in the CFL ground state,
which would be a trigger of the chiral condensate.
Actually, the coexistence phase of the chiral and diquark condensates at 
high baryon density 
has been recently reported based on the model-independent Ginzburg-Landau 
approach 
taking into account the instanton effects properly \cite{HTYB06, YTHB07}.
The important point there is that the instanton-induced interaction 
composed of the chiral and diquark condensates:
\beq
\label{eq:ext}
{\cal L}_{\rm ext} &=& \gamma {\rm Tr}[(d_R d_L^{\dag})(\bar q_{R} q_{L}) + 
{\rm h.c.}],
\eeq
acts an external field for the chiral condensate and
leads to a chirally broken crossover between the hadronic phase and the CSC 
phase.
Our result of the coexistence phase at high baryon density 
is totally consistent with this observation due to the same origin of
instantons.

It should be remarked that the chiral condensate in dense QCD is 
proportional to the instanton density
in Eq.~(\ref{eq:chiral}),
which is in contrast with the case of the QCD vacuum with $N_f \geq 2$
where the chiral condensate
is expected to behave as \cite{SS98}
\beq
\label{eq:chiral_vac}
\langle \bar q q \rangle_{\rm vac} \propto - \frac{n_{\rm 
inst}^{1/2}}{\rho}.
\eeq 
This difference can be understood as follows: 
The spontaneous breaking of chiral symmetry in the QCD vacuum is a 
collective phenomena 
caused by the effect of infinitely many instantons,
and the chiral condensate must be determined from self-consistent 
relations,
which finally results in Eq.~(\ref{eq:chiral_vac}) \cite{SS98}.
On the other hand, in the case of dense QCD, chiral symmetry is broken 
by a single instanton effect thanks to the presence of diquarks as shown in 
Eq.~(\ref{eq:ext}),
and it is anticipated that the chiral condensate is proportional 
to the number of instantons $N$ in a four-volume $V_4$, i.e., the instanton 
density $n_{\rm inst}$.

\section{Discussion and conclusion}
\label{sec:discussion}
In this paper, we have studied the properties of an instanton ensemble in 
three-flavor dense QCD which can be regarded as an instanton plasma
weakly interacting by exchanging the $\eta'$ mesons. 
Based on this description, we derive analytical formulas for 
the instanton density, the topological susceptibility and a 
dense version of the Witten-Veneziano relation.
We also explore the chiral phase transition induced by the instanton 
ensemble
in analogy with the Berezinskii-Kosterlitz-Thouless transition.
We generally show that the system of Coulomb interacting $D$-dimensional 
topological excitations 
exhibits a second order phase transition for $D=2$, and a crossover for $D 
\ge 3$ using the renormalization
group approach.
In particular, for $D=4$, the instanton ensemble always behaves as a 
screened and unpaired plasma, which
gives nonvanishing chiral condensate proportional to the instanton density 
at high baryon density regime of QCD.
Therefore, the coexistence phase of the chiral and diquark condensates 
is inevitably expected in dense QCD as suggested in Refs.~\cite{HTYB06, 
YTHB07}.

The discussion on the applicable domain of the instanton description 
introduced above is in order here.
Our treatment is based on the low-energy effective Lagrangian of 
the $\eta'$ meson, Eq.~(\ref{eq:eta'-Lagrangian}),
which is valid when two conditions on the $\eta'$ pole mass are satisfied:
(i) $m_{\eta'} \lesssim 2 \Delta$, and (ii) $m_{\eta'} \lesssim 
m_{\pi,K,\eta}$.
The condition (i) is required since, otherwise ($m_{\eta'}>2\Delta$), 
$\eta'$ would rapidly decay into a particle-hole pair 
and becomes unstable. 
Also the condition (ii) is necessary to assure that we have only to
focus on the low-energy effective Lagrangian of the light $\eta'$ meson.
When $\mu \gg \Lambda_{\rm QCD}$, 
$m_{\eta'} \ll 2\Delta$ as well as 
the inverse meson mass ordering, $m_{\eta'}<m_K<m_{\pi}<m_{\eta}$ 
follows due to $a \ll 1$, so that the conditions (i) and (ii) are satisfied 
(See Appendix \ref{sec:spectra}).
Moreover, we find the critical chemical potential $\mu_c$ as
$\mu_c \sim 10 \Lambda_{\rm QCD}$ for
$m_{ud}=5$-$10$ MeV, $m_s=150$ MeV and $\Lambda_{\rm QCD}=200$ MeV.

The extrapolation of the instanton-induced crossover obtained here 
to lower baryon density is a nontrivial question
which we cannot address within our treatment.
However, it might be reasonable to expect that the system of instantons 
behaves as a gas-like weakly-correlated or 
a liquid-like strongly-correlated plasma 
across the entire span of the density.
The instanton-induced crossover may have relevance 
to the continuity between hadronic phase and color 
superconductivity phase \cite{SW99, F04} and 
the spectral continuity of hadrons \cite{YTHB07, HTY08} 
from low to high baryon densities.
It would be also important to study 
how the confinement-deconfinement phase transition at finite baryon density 
is related to the changes in the behavior of an instanton ensemble 
\cite{TZ06}.

\acknowledgments
The author would like to thank T. Hatsuda for discussions,
helpful comments and careful reading of the manuscript.
He also thank K. Iida, R. Millo, S. Sasaki and M. Tachibana 
for useful conversations.
He is grateful to the ECT* program on ``Nuclear Matter under Extreme 
Conditions"
where a part of this work has been completed.
The hospitality of the
European Centre for Theoretical Studies in Nuclear
Physics and Related Areas (ECT*)
is gratefully acknowledged. 
The work is supported by the Japan Society for the Promotion of Science
for Young Scientists.

\appendix
\section{Mass spectra of meson excitations}
\label{sec:spectra}
The explicit inclusion of the ${\cal O}(M^2)$-term
does not change our discussion in a substantial way.
But it is rather essential to validate the low-energy effective Lagrangian 
of the $\eta'$ meson, 
Eq.~(\ref{eq:eta'-Lagrangian}), so that we can neglect other heavier meson 
excitations.
In the case of two light degenerate up and down quarks with a medium-heavy 
strange quark, 
flavor $SU(2)$ symmetry is respected but flavor $SU(3)$ symmetry is not. 
Then we find the masses of pions ($\pi^0$ and $\pi^{\pm}$) 
and kaons ($K^{\pm}$, $K^0$ and ${\bar K}^0$) to the order of ${\cal 
O}(M^2)$ as 
\cite{SS00, MT00, BBS00, S01}:
\beq
\label{eq:pi}
m_{\pi^{\pm}, \pi^0} &=& 
\left[\frac{2a}{f_{\pi}^2}m_{ud} + \frac{8C}{f_{\pi}^2}m_{ud} m_s 
\right]^{1/2}, \\
\label{eq:K}
m_{K^{\pm}, K^0, {\bar K}^0} &=& \mp \frac{m_s^2-m_{ud}^2}{2\mu} +
\left[\frac{a }{f_{\pi}^2} (m_{ud} + m_s) + \frac{4C}{ f_{\pi}^2}m_{ud} 
(m_s+m_{ud})\right]^{1/2},
\eeq
where the first term on the right hand side of Eq.~(\ref{eq:K}) is the 
effective modifications of the chemical potential 
due to the explicit breaking of the flavor $SU(3)$ symmetry \cite{BS02}.  
The coefficient $C$ and the pion decay constant $f_{\pi}$ 
have been determined from weak-coupling calculations at high density 
\cite{SS00}:
\beq
C&=&\frac{3\Delta^2}{4\pi^2}, \\
f_{\pi}^2&=&\frac{21-8\ln2}{18}\frac{\mu^2}{2\pi^2}.
\eeq
On the other hand, the neutral mesons, $\eta$ and $\eta'$, are unaffected 
by the effective
chemical potential. However, since $\eta'$ mixes with $\eta$, 
the diagonalization of the $2 \times 2$ mass matrix $m_{ab}^2$ $(a,b=0,8)$,
\beq
m_{00}^2 &=& \frac{8C}{3f_{\eta'}^2}m_{ud}(2m_s + m_{ud}), \nonumber \\
m_{08}^2 &=& \frac{8\sqrt2 C}{3f_{\eta'}f_{\pi}}m_{ud}(m_s - m_{ud}), \\
m_{88}^2 &=& \frac{8C}{3f_{\pi}^2}m_{ud}(m_s + 2 m_{ud}), \nonumber
\eeq
is necessary to obtain the genuine mass eigenvalues of $\eta'$. 
Also taking into account the instanton contribution to $\eta'$, 
Eq.~(\ref{eq:eta'-mass}), 
their masses finally turn out to be
\beq
\label{eq:eta'}
m_{\eta'} &=& 
\left[ \frac{2 a }{3f_{\eta'}^2}M +  \frac{24C}{2f_{\pi}^2 + f_{\eta'}^2} 
m_{ud}^2 \right]^{1/2}, \\
\label{eq:eta}
m_{\eta} &=& 
\left[\frac{a}{3f_{\pi}^2}(m_u + m_d + 4 m_s) 
+ \left(\frac{1}{f_{\pi}^2}+\frac{2}{f_{\eta'}^2} \right)
\frac{8C}{3}m_{ud}m_s 
\right.  \nonumber \\
& & \left.
+ \left(\frac{2}{f_{\pi}^2}+\frac{1}{f_{\eta'}^2}
 - \frac{9}{2f_{\pi}^2 + f_{\eta'}^2} \right) \frac{8C}{3}m_{ud}^2
\right]^{1/2}.
\eeq
 

\end{document}